# A Clinically Anchored Radiomics Dictionary for Explainable TI-RADS–Based Thyroid Nodule Classification in Ultrasound; Dictionary Version TU1.0

Mohammad Salmanpour[*,1,2,3], Shahram Taeb[4], Ali Fathi Jouzdani[3,5], Mohammad Ayazi[6], Siavash Hosseinpour Saffarian[7], Mehdi Maghsudi[3], Ilker Hacihaliloglu[2,8], Arman Rahmim[1,2]

[1]Department of Basic and Translational Research, BC Cancer Research Institute, Vancouver, BC, Canada
[2]Department of Radiology, University of British Columbia, Vancouver, BC, Canada
[3]Technological Virtual Collaboration Company (TECVICO CORP.), Vancouver, BC, Canada
[4]Department of Radiology, School of Paramedical Sciences, Guilan University of Medical Sciences, Rasht, Iran
[5]NAIRG, Department of Neuroscience, Hamadan University of Medical Sciences, Hamadan, Iran
[6]School of Medicine, Kermanshah University of Medical Science, Kermanshah, Iran
[7]College of Interdisciplinary Science and Technology, Faculty of Intelligent Systems, University of Tehran, Tehran, Iran
[8]Department of Medicine, University of British Columbia, Vancouver, BC, Canada

(*) Corresponding Author: Mohammad Salmanpour, PhD; msalman@bccrc.ca

## Abstract

**Purpose:** Artificial Intelligence-based radiomics models for thyroid ultrasound (US) often lack interpretability, limiting clinical trust. This study develops and validates an interpretable radiomic feature (RF) framework for thyroid-nodule classification by linking quantitative-US features to the Thyroid-Imaging-Reporting-and-Data-System (TI-RADS) semantic lexicon through a clinically grounded radiomics dictionary.

**Methods:** A radiomics dictionary was constructed to map TI-RADS categories, including composition, echogenicity, shape, margin, and echogenic foci, to Image Biomarker Standardization Initiative–compliant RFs extracted from two-dimensional-US images. Relationships were defined through expert consensus (four physicians, three physicists, one radiology expert, one biologist) and validated using Shapley-Additive-Explanations (SHAP). Three multicenter datasets were combined, yielding 5,542 nodules, from which 107-RFs were extracted using PyRadiomics and normalized with min–max scaling. Twenty-seven feature-selection methods were paired with twenty-five classifiers and evaluated using stratified five-fold cross-validation on 70% of the data, followed by testing on the remaining 30% for benign-versus-malignant nodule classification. Robust model selection employed a stability-aware composite scoring framework combining mean performance and variability across balanced accuracy, precision, recall, F1-score, and Receiver Operating Characteristic– Area Under the Curve (ROC-AUC).

**Results:** The dictionary enabled direct interpretation of radiomic signatures in TI-RADS terms. The Select-From-Model (logistic regression) plus Extra-Trees classifier achieved strong testing performance (ROC-AUC: 0.941±0.005). SHAP identified texture heterogeneity as the dominant malignancy signal, with Gray Level Run Length Matrix non-uniformity, intensity dispersion, and kurtosis aligning predictions with high-risk TI-RADS descriptors.

**Conclusion:** This study introduces an interpretable radiomics dictionary and stability-aware model selection framework, overcoming black-box limitations and enabling transparent thyroid nodule risk stratification from US.

## 1. Introduction

Thyroid nodules are discrete lesions within the thyroid gland and are among the most common findings on endocrine imaging, with prevalence increasing with age and the widespread use of ultrasound (US, all abbreviations are defined in Appendix Table A1) [1]. Although the thyroid nodules classification (TNC) is benign and clinically indolent, a small yet clinically significant proportion represents thyroid cancer (TC), which typically manifests as malignant nodules on imaging [2]. TC, the most common endocrine malignancy worldwide, has seen its incidence increase dramatically over the past several decades [1]. Consequently, accurate differentiation of TNC is a critical clinical challenge, as early detection of TC directly impacts diagnostic strategies, biopsy recommendations, and therapeutic decision-making. However, the substantial overlap in sonographic features between benign and malignant nodules limits the reliability of conventional US assessment, underscoring the need for accurate, reproducible risk stratification methods that minimize unnecessary interventions while ensuring timely identification of clinically relevant thyroid malignancies [3].

The US is the first-line imaging modality for TNC. It is widely available, cost-effective, radiation-free, and capable of providing high-resolution visualization of thyroid anatomy [4]. Conventional US assessment relies on qualitative features such as: Echogenicity, Margins and shape, Calcifications, and Composition (solid vs cystic). However, this visually driven approach is inherently subjective and dependent on operator expertise, limiting reproducibility and sensitivity to subtle imaging patterns. In this context, artificial intelligence (AI), and particularly the discipline of radiomics, has emerged as a transformative approach with the potential to objectify and refine TNC [5]. Radiomics is a known quantitative approach to medical imaging that facilitates the high-throughput extraction of a vast number of mathematical features from standard-of-care medical images [6]. These features, far exceeding the capacity of human visual perception, capture a deep and granular profile of a tumor's phenotype, quantifying its shape, intensity distribution, and intricate textural patterns [7] By converting images into high-dimensional, mineable data, radiomics can uncover subtle patterns of intratumoral heterogeneity that are directly linked to the underlying pathophysiology, such as gene expression patterns and cellular behavior, including proliferation [8]. When coupled with machine learning (ML) and deep learning algorithms, radiomics-based models have demonstrated strong predictive performance across multiple oncologic applications, including treatment response and patient survival [9] [10] [11] [12] [13] [14]. In thyroid imaging, this approach holds the promise of developing highly accurate, non-invasive biomarkers for malignancy, thereby reducing the need for unnecessary biopsies [5].

Despite this immense potential, a significant and persistent barrier to widespread clinical adoption and translation of these powerful AI-driven models is the inherent "black box" problem [15]. Many powerful ML algorithms operate in ways that are not transparent to the end-user. The abstract, mathematical nature of many radiomic features (RFs)—such as "Gray Level Co-occurrence Matrix Inverse Difference Moment" makes it especially hard for clinicians to understand the logic behind a model's prediction [16]. The inability to interpret these algorithms undermines trust and confidence among clinicians, as doctors are understandably hesitant to depend on algorithmic results that they cannot comprehend or validate [17]. For AI to be truly integrated into the clinical workflow, its outputs must be both precise, explainable, and consistent with established medical knowledge.

In response to this diagnostic uncertainty and the need to reduce unnecessary biopsies, a concerted effort was made to develop a standardized framework for US-based risk stratification [18]. To standardize the risk stratification of TNC and reduce inter-observer variability, the American College of Radiology (ACR) developed and disseminated the Thyroid Imaging Reporting and Data System (TI-RADS) [19]. TI-RADS offers a structured, evidence-based system that assigns points to TNC based on five key categories of suspicious US features: composition (e.g., cystic, solid), echogenicity (e.g., hypoechoic), shape (e.g., taller-than-wide), margin (e.g., irregular), and echogenic foci (e.g., punctate microcalcifications). The total point score assigns a nodule to a specific risk category (TR1-TR5), indicating whether to recommend a fine-needle aspiration (FNA) biopsy or imaging follow-up [20]. The widespread adoption of TI-RADS has played a crucial role in standardizing the TNC. However, it remains a primarily qualitative system that depends on human visual interpretation of these semantic features. As a result, it cannot wholly eliminate subjectivity and may miss some aspects of a tumor's biological complexity that quantitative methods could reveal [19].

A critical and unaddressed gap thus exists at the intersection of these two powerful but separate domains, the computational realm of radiomics and the established, semantic lexicon of TI-RADS [15]. A radiomics model might consider a high value for a texture feature, such as "Gray Level Non-Uniformity" (GLN), as a strong predictor of malignancy. However, it remains entirely unclear for a radiologist how this quantitative measure biologically or visually relates to a clinically recognized and trusted TI-RADS descriptor, such as an "ill-defined margin" or "marked hypo-echogenicity." This lack of a translational link is the primary obstacle preventing the rich, objective, and reproducible insights from radiomics from being easily understood, validated, and practically incorporated into the standardized clinical workflow established by TI-RADS.

In recent decades, several studies have explored radiomics dictionaries to establish relationships between RFs and standardized clinical scoring systems, such as Reporting and Data Systems (RADS) frameworks or World Health Organization (WHO) classifications. These approaches have been applied in prostate [16], breast [15], lung [21], and liver [17] cancers, where feature–semantic mappings have enhanced model interpretability and strengthened clinical relevance. However, a notable gap remains in TC, as no comprehensive and clinically grounded radiomics dictionary has yet been developed to link US-derived RFs with TI-RADS semantic descriptors. To address this critical translational gap, this study introduces the first Radiological and Biological Dictionary of RFs specifically for TC. This framework will clearly connect the main quantitative RFs from thyroid US images to the standardized visual semantic descriptors in the ACR TI-RADS lexicon. By establishing this organized and understandable link, we aim to clarify the "black box" of radiomics for TC evaluation. This study introduces the first Radiological and Biological Dictionary of RFs in TC by establishing robust statistical and visual correlations between quantitative RFs and the standardized visual semantic descriptors. So, this study aims to develop and validate a clinically anchored Radiological and Biological Dictionary that explicitly maps Image Biomarker Standardization Initiative (IBSI)-compliant RFs to ACR TI-RADS semantic descriptors, thereby enabling explainable, reproducible, and clinically interpretable ML–based TNC.

## 2. Materials and Methods

### 2.1. Development of a Clinically Informed Feature Interpretation Dictionary

Figure 1 illustrates the workflow of this study, from the development of the radiomics dictionary to an example of its application in TC risk stratification using TI-RADS–guided radiomic interpretation. The ACR TI-RADS is a standardized point-based framework for stratifying TNC risk using five US feature categories: composition, echogenicity, shape, margin, and echogenic foci. However, TI-RADS scoring relies on qualitative visual assessment and is susceptible to subjectivity and inter-observer variability. To address this limitation, we developed a Radiological and Biological Dictionary of RFs that quantitatively maps each TI-RADS descriptor to its underlying radiomic signature. This framework translates semantic sonographic impressions into objective, reproducible, and clinically interpretable biomarkers by linking TI-RADS point assignments to algorithmically derived RFs. Building on prior work in breast oncology (BM1.0) [15], our dictionary framework (TU1.0) provides a systematic translation layer between established radiological signs and quantitative imaging features, enabling interpretable integration of radiomics within clinical TI-RADS reporting.

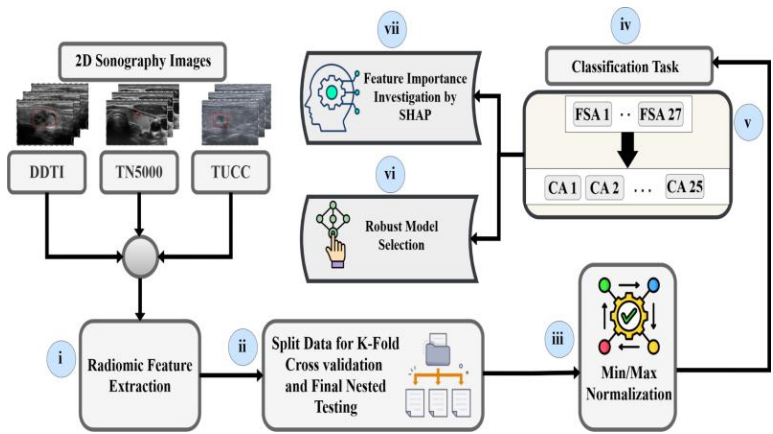

**Fig. 1.** Overview of the proposed radiomics-based ML pipeline for thyroid US image classification. The workflow consists of multiple sequential stages: (i) RFs extraction using standardized PyRadiomics descriptors; (ii) splitting the data for CV and final testing; (iv) applying min–max normalization using parameters derived from training folds; (iv) implementing a supervised ML approach with fivefold CV; (v) applying FSA and benchmarking multiple CA; (vi) evaluating model performance using standard metrics; and (vi) investigating feature importance and model interpretability through SHAP analysis. This integrated pipeline ensures robust model development, performance assessment, and explainability across diverse thyroid US datasets. Feature selection algorithms (FSA), Shapley Additive exPlanations (SHAP), Classification Algorithms (CA), machine learning (ML), ultrasound (US), radiomic features (RFs), cross-validation (CV).

The reference feature set comprised IBSI-defined RF classes, including first order (FO), intensity distribution, Shape-based metrics (lesion morphology), and multiple texture families such as gray level co-occurrence matrix (GLCM) (spatial gray-level relationships), Gray Level Size Zone Matrix (GLSZM) (zone-size homogeneity), gray level run length matrix (GLRLM) (run-length patterns), neighboring gray-tone difference matrix (NGTDM) (local intensity variation), and gray level dependence matrix (GLDM). To establish conceptual links between quantitative RFs and qualitative TI-RADS descriptors (e.g., "taller-than-wide, "very hypoechoic," punctate echogenic foci," "lobulated/irregular margin"), each feature was systematically evaluated for interpretive relevance to the underlying visual characteristic. This process generated a clinically informed feature-interpretation dictionary that translates descriptive radiological terminology into computable imaging biomarkers, thereby improving both the explainability of AI-driven models and the clinical interpretability of radiomic outputs. Dictionary validation and refinement were performed through an expert-driven review process, in which domain experts, such as four physicians, three physicists, one radiology expert, and one biologist, first assessed global relationships between TI-RADS semantic attributes and the full RFs candidate set, then iteratively refined associations to identify the most semantically faithful RFs for each descriptor.

**2.2 Explainable Machine Learning Classification Task**

**2. 2.1. Patient Dataset**

Large multicenter datasets, including 5,829 2D-US images and their corresponding annotations, were collected from three publicly available databases: Thyroid US Cine-clip (TUCC), TN5000, and Digital Database Thyroid Image (DDTI). The TUCC dataset comprises 192 US images with biopsy-confirmed thyroid nodules [22], TN5000 dataset includes 4,974 US cases [23], and the DDTI dataset encompasses 376 patients with histopathologically verified nodules [24]. Histopathological diagnoses were available for all cases across the three cohorts. In the TUCC and TN5000 datasets, diagnostic labels were binary, with "1" denoting malignant and "0" denoting benign nodules. The DDTI dataset employed a multi-class TI-RADS-inspired labeling system (1, 2, 3, 4a, 4b, 4c, and 5), representing increasing levels of malignancy risk from unequivocally benign (Label 1) to highly suspicious for malignancy (Label 5). Since the labels were imbalanced, for consistency and to enable binary classification, labels 1, 2, and 3 were grouped as benign (Class 0), while labels 4a, 4b, 4c, and 5 were grouped as malignant (Class 1). When multiple nodules were present in a single patient, each nodule was treated as an independent case to preserve statistical independence. Cases lacking segmentation masks or with incomplete annotation were excluded from all datasets to ensure robust feature extraction (n=287). Table 1 presents a summary of the three datasets. During the US examinations, machines comparable to widely used clinical models reported in the literature, such as the GE Logiq E9 or S7 series, were employed. These systems typically operate within frequency ranges of 5–12 MHz or 8–15 MHz, acquiring diagnostic images through a transducer probe guided by the sonographer.

**Table 1.** Summary of the datasets used in this study, including the total number of cases and the distribution of benign and malignant nodules.

| Dataset | Benign (0) | Malignant (1) | FNA Biopsy | Total Number of Images | Excluded | Min Size | max size |
|---|---|---|---|---|---|---|---|
| TUCC | 175 | 17 | Done | 192 | 0 | - | - |
| TN5000 | 1,426 | 3,548 | Done | 5,000 | 26 | (439, 368) | (919, 646) |
| DDTI | 139 | 237 | Done | 637 | 261 | (245, 360) | (560, 360) |

**2.2.2. Classification Machine Learning Task**

As illustrated in Figure 1, the proposed pipeline presents a comprehensive and clinically aligned framework for developing robust, interpretable ML models based on RFs extracted from 2D thyroid US images. The workflow encompasses rigorous image preprocessing, targeted dimensionality reduction via feature selection algorithms (FSA), systematic classifier benchmarking, and stringent validation of supervised learning models using cross-validation (CV) and final nested testing. In the interpretation phase, the selected RFs are analyzed individually and in groups to understand their biological relevance and diagnostic contribution, rather than being transformed through attribute extraction algorithms. This end-to-end pipeline is meticulously designed to optimize feature representation while retaining interpretability, enhance diagnostic generalizability, and ultimately improve the clinical utility of ML classifiers in differentiating benign from malignant TNC, thereby supporting trustworthy AI integration into TC care.

**Radiomics Feature Extraction**. Using a fixed PyRadiomics configuration, 107 RFs were extracted from each image. For cross-dataset analysis, feature alignment was performed by retaining only those features available across all three datasets, resulting in a final common set of 93 RFs used for subsequent modeling [25]. The complete list of extracted for each dataset and excluded features is provided in the Supplemental File 1 (Sheet 1).

**Splitting Data**. Following feature extraction, the TUCC, TN5000, and DDTI datasets were first combined and then partitioned into training and testing components. To construct a final test set, 30% of the cases from each dataset were randomly sampled while enforcing an equal number of benign and malignant nodules to mitigate class imbalance. The remaining 70% of cases, drawn proportionally from all three datasets, were used for model development under a five-fold CV scheme. In total, the training set consisted of 3,880 cases, and the final test set contained 1,662 cases, precisely balanced between 831 benign and 831 malignant nodules. This design ensured that model training utilized the full diversity of the multi-institutional data while maintaining an unbiased and diagnostically balanced final evaluation cohort.

**Min/Max Normalization**. To ensure methodological rigor and prevent data leakage, the normalization parameters (i.e., minimum and maximum values for min–max scaling) were computed exclusively from the training folds—stratified into four subsets via 5-fold CV—and subsequently applied to the validation and final test sets during inference, ensuring that no information from evaluation data influenced the scaling process in the context of 2D thyroid US-derived radiomics for AI-driven malignancy stratification.

**Classification Task**. The dataset was stratified into five balanced folds for stratified five-fold CV, with 80% of the 3,880 cases used for training and 20% for validation in each iteration. This procedure was repeated to minimize sampling bias and ensure robust performance estimation. The best-performing models were subsequently evaluated on an independent 30% held-out multi-center test cohort (1,662 cases) to assess cross-institutional generalizability. Model performance was quantified using Accuracy, Precision, Recall, F1-score, Specificity, and Receiver Operating Characteristic– Area Under the Curve (ROC-AUC), reported as mean ± standard deviation (SD) across the five folds and separately for the test set. Final model selection was based on balanced CV results and validated on unseen data to ensure clinical reliability and reproducibility.

**Dimensionality Reduction**. To address the high dimensionality of RFs extracted from 2D thyroid US and mitigate overfitting, the pipeline incorporates 27 FSAs. FSAs identify a subset of the original, clinically interpretable RFs; in this study, each FSA was configured to select the top 10 most relevant features. The 27 FSAs span multiple methodological families to ensure robust identification of informative and non-redundant RFs. Filter-based approaches, including Chi-Square Test, Chi-squared P-value, Correlation Coefficient, Mutual Information, Mutual Information Selection, Mutual Information Gain Ratio, Information Gain, Variance (V) Threshold, and ReliefF, provide rapid, model-agnostic ranking based on statistical dependency, V structure, or neighborhood-based relevance estimation. Multiple-comparison control strategies, such as False Discovery Rate (FDR) and Family-wise error (FWE), enhance statistical robustness during significance-based selection. Wrapper methods, including Recursive Feature Elimination (RFE), Sequential Feature Selection, and Univariate Feature Selection, iteratively optimize feature subsets using classifier feedback to improve predictive performance. Embedded methods, such as Select-From-Model with Logistic Regression (SMLR), LASSO, Elastic Net, Embedded Elastic Net, and Stability LASSO, integrate feature selection directly within model fitting through regularization and coefficient shrinkage to promote sparsity and reduce overfitting. Ensemble-based approaches, including Random Forest Importance (RFI), Extra Trees Importance, and Permutation Importance, capture nonlinear dependencies and interaction effects reflective of thyroid US heterogeneity. Additionally, multicollinearity-aware and structure-guided strategies such as Variance Inflation Factor (VIF) selection and principal component analysis (PCA)-based methods, including PCA loadings-based and PCA dictionary approaches, assist in stabilizing the selection of correlated RF groups. Collectively, these complementary methods support effective dimensionality reduction while preserving clinically interpretable radiomic variables.

**Classification Algorithms**. Each reduced feature representation—FSA-selected feature subsets-derived latent embeddings—was rigorously evaluated using a diverse ensemble of 25 ML classifiers spanning a wide range of interpretability, complexity, and learning paradigms. Tree-based classifiers included Decision Trees, Random Forest (RandF), Extra Trees Classifier (ETC), Gradient Boosting (GB), AdaBoost Classifier (ABC), and HistGradient Boosting (HGB), all of which leverage ensemble learning to mitigate overfitting, reduce V, and enhance generalization performance—particularly critical in small-sample, high-dimensional radiomics settings. Meta-ensemble strategies, such as Stacking, Voting Classifiers (hard and soft), and Bagging Classifiers (BaC), further improved predictive robustness by strategically aggregating predictions from heterogeneous base learners to exploit complementary strengths. Support Vector Machines (SVM) with linear, polynomial, and radial basis function (RBF) kernels were employed to effectively model both linear and nonlinear decision boundaries in the sonographic feature space. The k-Nearest Neighbors (KNN) algorithm provided a nonparametric, instance-based approach grounded in local similarity metrics derived from grayscale and textural patterns in 2D thyroid US images. Several Naive Bayes variants—Gaussian, Bernoulli, and Complement—were evaluated for their computational efficiency and baseline performance in probabilistic classification, especially useful given the often-limited sample sizes in thyroid imaging cohorts. Neural network–based Multi-Layer Perceptron (MLP) models were configured to capture complex, nonlinear interactions among sonographic RFs, while gradient-boosted frameworks, including Light Gradient Boosting Machine (LGBM) and XGBoost (XGB), delivered state-of-the-art performance through efficient gradient optimization, built-in regularization, and intrinsic feature importance quantification. Additional classifiers included Linear Discriminant Analysis (LDA) for dimensionality reduction and class separation, Nearest Centroid for prototype-based classification, Decision Stump as a minimal interpretable baseline, Dummy Classifier (DUC) for chance-level benchmarking, Gaussian Process Classifier (GPC) for uncertainty-aware prediction, and Stochastic Gradient Descent Classifier (SGDC) for scalable linear modeling. All Classification Algorithms (CAs) were optimized using fivefold CV combined with exhaustive grid search for hyperparameter tuning, ensuring robust model selection while minimizing data leakage and overfitting—a critical consideration in thyroid US–based radiomics where inter-scanner and inter-operator variability can significantly affect feature stability.

**Robust Model Selection.** We implemented a statistically rigorous model evaluation and selection pipeline for learning tasks in multicenter studies, designed to compare all FSA-classifier combinations while preventing information leakage fairly [26]. Consistent with the data-partitioning strategy detailed in iii) Splitting Data, model scoring and ranking were performed exclusively using 5-fold CV results from the 70% model development cohort, while the 30% final test set remained fully blinded and played no role in model selection. We normalize the metric means (Equation 1) and SD (Equation 2) across all models:

CV Evaluation: As established in iii) Splitting Data, following feature extraction, the TUCC, TN5000, and DDTI datasets were merged and then stratified into training and testing partitions. Specifically:
- 30% of cases from each dataset were randomly sampled to form a final test set, enforcing parity between benign and malignant nodules to mitigate class imbalance.
- The remaining 70% of cases, proportionally drawn from all datasets, formed the model development (training/validation) cohort, which was used to assess model configurations via 5-fold CV.

Within this single rotation, for each model and metric i:
- Mean performance across folds: $\mu_i$ = mean over 5 fold CV
- Model stability: $\sigma_i$ = SD across 5 fold CV

Therefore, each model generated performance metrics during CV: Accuracy, F1 Score, Precision, Recall, and ROC-AUC. For each model–metric pair (from the 70% model development cohort):
- $\mu_i$ = mean of metric i over 5 CV folds
- $\sigma_i$ = SD of metric i over 5 CV folds

Normalization and Stability Scoring: To enable fair ranking across all models, we applied min–max normalization computed globally across all models per metric:

Normalized mean score (Eq. 1):

$$\widehat{M}_i = \frac{\mu_i - min\ (M_i)}{max\ (M_i) - min\ (M_i)} \quad (1)$$

Normalized SD (Eq. 2):

$$\hat{S}_i = \frac{\sigma_i - min\ (S_i)}{max\ (S_i) - min\ (S_i)} \quad (2)$$

Stability transformation (higher = more stable) (Eq. 3)

$$Stability_i = 1 - \hat{S}_i \quad (3)$$

Final Composite Model Score: Model scores aggregated performance and stability equally across the single 5-fold CV (Eq. 4):

$$Final\ Score = \frac{1}{20} \sum_{i=1}^{5} (\widehat{M}_i + Stability_i) \quad (4)$$

- This reflects 10 metrics (including averages and SDs) × 2 components = 20 score terms.
- The final score is constrained to [0, 1] with equal weighting of mean performance and stability, and no contribution from the final test set.
- 5 internal mean values (Accuracy, F1, Precision, Recall, ROC-AUC)
- 5 internal SD values, subsequently converted to stability estimates
  (Total internal estimates per model = 10, all from CV)

**Shapley Additive Explanations -Based Development of a Data-Driven Radiomics Interpretation Dictionary.** The objective of Shapley Additive Explanations (SHAP) is to interpret model predictions by quantifying how each RF contributes to classifying a thyroid nodule as malignant or benign. SHAP is based on Shapley values from cooperative game theory, where each RF is treated as a "player" in a coalition, and the model's predicted malignancy probability represents the "payout" that must be fairly distributed among all features [27]. Shapley values estimate feature importance by computing the average marginal contribution of each RF across all possible feature combinations. This allows SHAP to capture not only individual feature effects but also interactions between features that may jointly influence malignancy risk. SHAP represents the explanation as an additive feature attribution model (Eq. 5):

$$g(z') = \phi_0 + \sum_{i=1}^{M} \phi_i z'_i \quad (5)$$

where $g(z')$ is the explanation model, $z'$ is a binary coalition vector indicating whether a feature is present (1) or absent (0), $M$ is the total number of RFs, and $\phi_i$ is the Shapley value contribution of the feature $i$. For the nodule instance being explained, all features are present ($z' = 1$), and the prediction can be written as (Eq. 6):

$$f(x) = \phi_0 + \sum_{i=1}^{M} \phi_i \quad (6)$$

Here, $\phi_0$ represents the baseline malignancy prediction (expected model output), while each $\phi_i$ indicates how much a specific RF shifts the prediction toward malignancy or benignity. Positive SHAP values increase the predicted malignancy risk, whereas negative values support a benign classification. Because SHAP satisfies key theoretical properties such as efficiency, symmetry, and additivity, it provides reliable and consistent explanations. This makes SHAP particularly valuable for identifying radiomic biomarkers that align model decisions with clinically recognized malignant and benign US patterns, thereby supporting transparent TC risk stratification.

# 3. Results

## 3.1. Association Between TI-RADS Scoring and Descriptor-Based Features

Figure 2 summarizes the complete ACR TI-RADS scoring workflow as a sequential, point-based risk stratification flowchart. Assessment begins with Step 1 (Composition): nodules categorized as cystic or spongiform receive 0 points and are immediately assigned TR1 (benign), ending the evaluation. Nodules that are mixed cystic and solid (1 point) or solid/almost completely solid (2 points) proceed to the remaining criteria. Step 2 (Echogenicity) assigns 0–3 points, from anechoic (0) to very hypoechoic (3). Step 3 (Shape) assigns 0 points for wider-than-tall orientation and 3 points for the high-risk taller-than-wide orientation. Step 4 (Margin) assigns 0 points for smooth margins, while suspicious patterns, including ill-defined (2 points) and lobulated/irregular (3 points), increase the score. Step 5 (Echogenic Foci) assigns 3 points for punctate echogenic foci, 1 point for macrocalcifications, and 0 points for none or large comet-tail artifacts. The points from all five steps are then summed to yield a total score, which determines the final TI-RADS category, ranging from TR1 (0 points, benign) to TR5 (>7 points, highly suspicious), and informs clinical management, including surveillance recommendations and FNA thresholds. A comprehensive explanation of the descriptors is provided in Supplemental File 1 (Sheet 2).

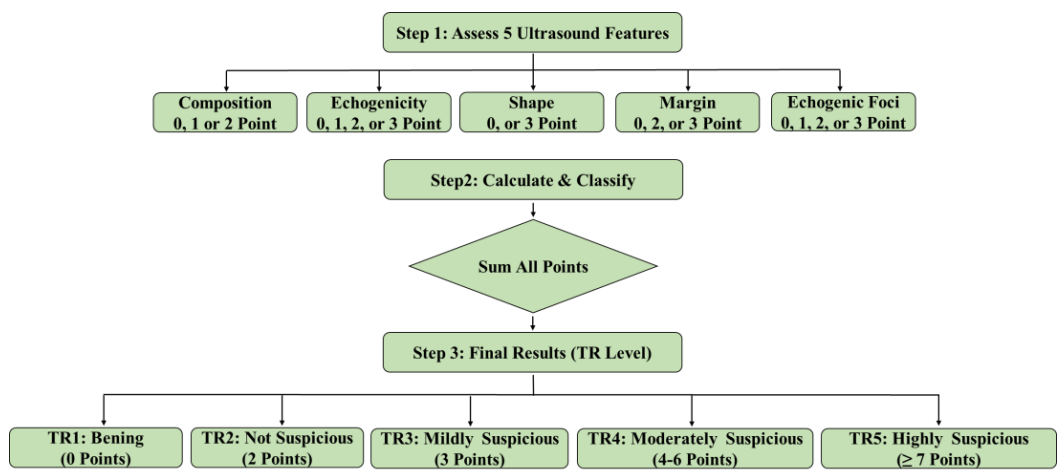

**Fig. 2.** The workflow of the ACR TI-RADS assessment system for TNC, illustrating the three-step process: Step 1 (i) evaluation of five US features (composition, echogenicity, shape, margin, echogenic foci) with assigned point values; Step 2 (ii) summation of points; and Step 3 (iii) classification into one of five risk categories (TR1–TR5), guiding clinical management based on malignancy suspicion. American College of Radiology-Thyroid Image Reporting and Data System (ACR TI-RADS), Thyroid Nodules Classification (TNC), Ultrasound (US).

## 3.2. Mapping TI-RADS Semantic Descriptors to RFs

This study results in the creation of the first comprehensive Radiological and Biological Dictionary of RFs specifically designed for US imaging of TC. These visual semantic features describe the key US characteristics used in thyroid nodule risk stratification, including composition, echogenicity, shape, margin characteristics, and echogenic foci, which together help determine the likelihood of malignancy. Detailed definitions and examples of each semantic descriptor are provided in Supplementary File 1 (Sheet 3). The dictionary provides a systematic framework that quantitatively links abstract RFs to the five established semantic categories (Cat) of the ACR TI-RADS. Our findings are organized sequentially around these five clinical categories, offering a clear and understandable connection between quantitative and qualitative TNC. A comprehensive mapping of TI-RADS semantic descriptors to RFs is provided in Supplemental File 1 (Sheets 4 and 5).

**Cat 1. Nodule Composition.** Nodule composition, which assesses the solid and cystic components of a nodule, is a crucial element of the TI-RADS classification. Our dictionary establishes a strong connection between these visual patterns and specific RFs that evaluate shape regularity and textural consistency.

(i) Cystic / Spongiform (Low Risk):

- Radiological Interpretation: These nodules, often linked to benign processes, exhibit highly regular, non-infiltrative growth patterns. This manifests as smooth, uniform shapes on 2D-US, reflecting their low malignant potential.
- Mapped RFs:
  - Circularity and Compactness from the Shape category: We observed a strong link between a benign composition and high values for shape regularity features. Circularity and Compactness quantitatively measure how closely a nodule's contour approximates a perfect circle on US images. The uniform, non-aggressive growth of cystic and spongiform nodules results in high scores for these features, providing a mathematical signature of their "round" and regular appearance.
  - Elongation from the Shape category: This feature, which displays the nodule's aspect ratio, was found to be low in cystic and spongiform nodules. This quantitatively confirms their tendency to be isodiametric (wider-than-tall), a pattern associated with benignity.

(ii) Mixed Cystic & Solid / Solid (Increased Risk)
- Radiological Interpretation: The presence of solid tissue introduces structural heterogeneity, which serves as a key indicator of higher malignant potential in TNC. This complexity is reflected in the distribution of echogenicity (pixel intensities) within the nodule on sonographic imaging.
- Mapped RFs:
  - Energy (E) and Entropy from the FO category: These two features serve as quantitative markers of tissue uniformity. A simple cyst that appears uniformly anechoic (hypoechoic/dark) has a straightforward intensity histogram, resulting in high E and low Entropy. Our analysis shows that as the nodule's composition becomes more solid and heterogeneous, the pixel distribution becomes more complex, causing a notable decrease in E and an increase in Entropy. These features provide reliable quantitative biomarkers for tissue complexity and solidity.

**Cat 2. Nodule Echogenicity**. Echogenicity is a critical factor in TI-RADS risk stratification. This visual characteristic directly reflects the nodule's internal cellular density and microstructure. Our dictionary offers a quantitative approach to interpret this qualitative feature by linking it to specific FO and texture-based RFs derived from 2D sonographic images.

(i) Hyperechoic / Isoechoic (Low to Mild Risk)
- Radiological Interpretation: Nodules that are equally echogenic to or brighter than the surrounding thyroid parenchyma are often associated with benign conditions, such as hyperplastic or colloid nodules. Their increased echogenicity reflects well-differentiated follicular structures that effectively reflect US waves.
- Mapped RFs:
  - Mean from the FO category: The mean intensity value, as the most direct measure of pixel echogenicity, serves as a reliable quantitative indicator. We believe that a high mean value corresponds to hyperechoic and isoechoic nodules.
  - Large Dependence High Gray-Level Emphasis (LDHGLE) from the GLDM category: This texture feature quantifies the extent of large, connected areas of high-echogenicity pixels. A high LDHGLE value indicates a coarse, uniformly echogenic texture, which is a hallmark of a hyperechoic nodule.

(ii) Hypoechoic / Very Hypoechoic (Moderate to High Risk)
- Radiological Interpretation: A nodule that appears darker than the surrounding thyroid parenchyma (hypoechoic) or the adjacent strap muscles (very hypoechoic) is a strong indicator of malignancy. This reduced echogenicity often reflects a high density of tightly packed neoplastic cells with minimal colloid, creating an architecture that scatters or absorbs US waves rather than reflecting them.
- Mapped RFs:
  - Mean from the FO category: A low mean pixel intensity is the primary quantitative indicator of a hypoechoic nodule, directly capturing its overall hypoechoic appearance.

- Low Gray-Level Run Emphasis (LGRE) from the GLRLM category: This feature specifically assesses the dominance of hypoechoic textural patterns. A high LGRE value indicates that the nodule's texture contains many consecutive runs of low-echogenicity pixels, providing a robust and specific quantitative validation of its hypoechoic nature.
- Coarseness from the NGTDM category: This feature quantifies the spatial rate of change in echogenicity. Malignant, hypoechoic nodules often exhibit a more intricate and finer internal texture due to cellular disarray. This results in abrupt intensity changes over short distances, leading to a low Coarseness value (i.e., a finer, more complex texture).

**Cat 3. Nodule Shape**

(i) Wider-than-Tall (Low Risk)
- Description: This describes a nodule with a non-aggressive, horizontal growth pattern, where the transverse dimension is greater than or equal to the anteroposterior (AP) dimension.
- Mapped RFs:
- Elongation from the Shape category (or Aspect Ratio): "To quantitatively capture this, the feature must be defined as the ratio of the AP Axis to the Transverse Axis. A low value ($\leq 1$) for this specific ratio provides a direct quantitative measure for this low-risk 'wider-than-tall' characteristic."

(ii) Taller-than-Wide (High Risk)
- Description: This describes a nodule with an aggressive, vertical growth pattern that disregards normal tissue planes. It is defined as the AP dimension being greater than the transverse dimension.
- Mapped RFs:
- Elongation from the Shape category (or Aspect Ratio): Using the identical feature definition (ratio of the AP Axis to the Transverse Axis), a high value ($> 1$) serves as the direct mathematical signature for a "taller-than-wide" nodule. This provides a clear, quantitative, and unambiguous flag for this critical high-risk finding.

Note on Dimensionality: While the standardized IBSI feature set includes Flatness, this feature depends on the Least Axis Length, which is mathematically undefined in 2D planar imaging. Therefore, for 2D thyroid US, Elongation remains the primary geometric descriptor for assessing the 'Taller-than-Wide' versus 'Wider-than-Tall' characteristic.

**Cat 4. Quantitative Correlates of Nodule Margin**. The characterization of a nodule's margin is a crucial part of risk assessment because it reveals how the nodule interacts with the surrounding thyroid tissue. A well-circumscribed, smooth margin typically indicates benign, contained growth, whereas an irregular, lobulated, spiculated, or ill-defined margin strongly suggests infiltrative malignant behavior. Our dictionary provides objective, quantitative measures to describe these boundary features in 2D-US images, bridging visual semantic descriptors from clinical reporting systems (e.g., TI-RADS) with reproducible RFs.

(i) Smooth (Low Risk)
- Radiological Interpretation:
- A smooth, well-circumscribed margin reflects a non-invasive growth pattern, wherein the nodule displaces rather than invades adjacent thyroid parenchyma. This manifests on US as a crisp, uninterruptedhypoechoic–hyperechoic interface with minimal acoustic shadowing or edge irregularity.
- Mapped RFs:
- Gradient Magnitude from the Gradient-Based category: This feature quantifies the rate of intensity change across the nodule boundary. A smooth, well-defined margin produces a steep, uniform intensity gradient, yielding high Gradient Magnitude values that objectively confirm sharp demarcation.
- Edge Sharpness from the Gradient-Based category: Complementary to Gradient Magnitude, this metric evaluates transition abruptness at the lesion edge. A smooth contour generates high Edge Sharpness, reflecting a clean, unambiguous interface on 2D-US.

  (ii) Irregular / Lobulated / Spiculated / Ill-defined (High Risk)

- **Radiological Interpretation:** These high-risk margin descriptors are hallmark signs of malignancy in TNC. An irregular or spiculated margin indicates direct tumor infiltration into perinodular tissue, often with microcalcifications or fibrous extensions. A lobulated contour suggests multicentric growth, while an ill-defined margin presents as a blurred, infiltrative halo, obscuring the nodule–parenchyma junction and complicating size assessment.
- **Mapped RFs:**
  - Gradient Magnitude and Edge Sharpness from the Gradient-Based category: In contrast to smooth margins, infiltrative or ill-defined boundaries produce gradual, heterogeneous intensity transitions, resulting in low Gradient Magnitude and low Edge Sharpness values—quantitative surrogates for boundary blurring and invasiveness.
  - Contrast and Dissimilarity from the GLCM category: When computed over a peri-nodular region of interest (ROI) encompassing the margin, these texture features capture local intensity variation. Spiculated or irregular margins induce pronounced pixel-level heterogeneity along the border, manifesting as elevated Contrast and Dissimilarity, thereby serving as robust mathematical proxies for margin irregularity and infiltrative potential in 2D-US imaging.

**Cat 5. Quantitative Correlates of Echogenic Foci**

(i) Punctate Echogenic Foci (High Risk)
- **Radiological Interpretation:** Small (<1 mm), discrete, dot-like hyperechoic foci without posterior acoustic shadowing, strongly associated with psammoma bodies in papillary thyroid carcinoma. Their presence confers one of the highest individual positive predictive values for malignancy in sonographic risk stratification (ACR TI-RADS Lexicon: "punctate echogenic foci"; European TI-RADS (EU-TI-RADS): "microcalcifications").
- **Mapped RFs:**
  - Kurtosis (Ku) from the FO category (FO_Ku): Punctate echogenic foci manifest as sparse, intensely hyperechoic pixels within a predominantly hypoechoic or isoechoic nodule matrix. This generates a leptokurtic intensity histogram with prominent heavy tails, reflected quantitatively by elevated Ku (>3.0). High Ku thus serves as a robust, reproducible surrogate for detecting these high-risk foci across US platforms.
  - The Ku feature, which is a FO statistical measure, tells you how peaked the intensity histogram is in the ROI. A histogram with high Ku values (over 3.0) has outliers, which means that the image has small, bright echogenic foci. This method works well for finding punctate hyperechoic foci, which are microcalcifications that radiologists have found, when used with the Laplacian of Gaussian (LoG) filter. The LoG filter makes it easier to find these structures, which lets you see fine-scale features. The average intensity of LoG responses in the ROI is a very specific biomarker for psammoma bodies that are linked to PTC.

(ii) Macrocalcifications / Peripheral (Rim) Calcifications (Variable Risk)
- **Radiological Interpretation:** Coarse (>2 mm), confluent hyperechoic regions or continuous curvilinear eggshell-like rim calcification. Macrocalcifications reflect chronic dystrophic change and occur in both benign (e.g., multinodular goiter) and malignant nodules. Interrupted or irregular rim calcification raises concern; complete, thin, uniform rims are typically benign.
- **Mapped RFs**
  - LDHGLE from the GLDM category: This feature quantifies the joint occurrence of high-intensity voxels over large spatial dependencies. Elevated LDHGLE (>75th percentile) reliably captures extensive, bright, contiguous calcified zones, providing mathematical fidelity to macrocalcifications and thick rim patterns. Integration with shape metrics (e.g., Surface Area to Volume Ratio) further distinguishes complete vs. disrupted rims, enhancing risk stratification.

(ii) Large Comet-Tail Artifacts (Reverberation) (Benign Indicator)
- **Radiological Interpretation:** V-shaped, tapering hyperechoic streaks extending posteriorly from a small, bright focus, pathognomonic for colloid crystals in benign cystic or macrofollicular nodules. Almost exclusively associated with benignity (ACR TI-RADS: "comet-tail artifact"; negative predictive value approaches 100% when multiple and classic).
- Small Dependence Low Gray-Level Emphasis (SDLGLE) from the GLDM category
  - Comet-tail artifacts comprise a compact, bright initiator (colloid) followed by a linear trail of progressively attenuating echoes. The SDLGLE feature from the GLDM category is specifically designed to measure small-scale joint occurrences of low-intensity voxels adjacent to high-intensity seeds. A high value (>0.7) for this feature in the immediate posterior ROI quantitatively encodes this reverberation tail pattern, serving as a highly specific benign biomarker. Furthermore, when combined with Minimum from the FO category (reflecting low tail intensity) and Cluster Shade from the GLCM category (indicating directional asymmetry), this feature reinforces confident benign classification. Table 2 summarizes the established relationships between RFs and TI-RADS semantic features.

**Table 2**. TI-RADS Semantic-to-Radiomic Mapping and Top Model Performance Evaluation; Thyroid Image Reporting and Data System (TI-RADS), gray level co-occurrence matrix (GLCM), gray level run length matrix (GLRLM), neighboring gray-tone difference matrix (NGTDM), gray-level dependence matrix (GLDM), Small Dependence Low Gray-Level Emphasis from the GLDM category (SDLGLE). Low Gray-Level Run Emphasis (LGRE), Kurtosis (Ku).

| TI-RADS Category | Semantic Descriptor | Risk Level | Mapped Radiomic Feature(s) | Quantitative Signature (Interpretation) |
|---|---|---|---|---|
| **Composition** | Cystic / Spongiform | Low | **Shape**: Circularity, Compactness **Shape**: Elongation | **High Value**: Indicates round/regular shape. **Low Value**: Indicates isodiametric shape. |
| | Mixed / Solid | Increased | **FO**: Entropy | **Low Value**: Indicates heterogeneity. **High Value**: Indicates complexity |
| **Echogenicity** | Hyperechoic / Isoechoic | Low / Mild | **FO**: Mean **GLDM**: LDHGLE | **High Value**: Indicates high average brightness. **High Value**: Indicates coarse, bright texture. |
| | Hypoechoic / Very Hypoechoic | Moderate. / High | **FO**: Mean **GLRLM**: LGRE, **NGTDM**: Coarseness | **Low Value**: Indicates overall darkness. **High Value**: Indicates dominance of dark texture runs. **Low Value**: Indicates fine, complex texture. |
| **Shape** | Wider-than-Tall | Low | **Shape**: Elongation (Aspect Ratio) | **Low Value** (≤ 1): Transverse axis ≥ AP dimension. |
| | Taller-than-Wide | High | **Shape**: Elongation (Aspect Ratio) | **High Value** (> 1): AP axis > Transverse axis |
| **Margin** | Smooth | Low | **Gradient**: Gradient Magnitude. **Gradient**: Edge Sharpness | **High Value:** Indicates steep intensity change at border. **High Value**: Indicates a distinct interface. |
| | Irregular / Ill-defined | High | **Gradient**: Gradient Magnitude. **GLCM**: Contrast, Dissimilarity | **Low Value**: Indicates gradual/blurred transition. **High Value**: Indicates boundary heterogeneity. |
| **Echogenic Foci** | Punctate Echogenic Foci | High | **FO**: Ku. | **High Value** (> 3.0): Indicates "peaked" histogram (outliers). **High Response**: Detects small bright blobs. |
| | Macrocalcifications | Variable | **GLDM**: LDHGLE | **High Value**: Indicates large, bright contiguous zones. |
| | Large Comet-Tail Artifacts | Benign | **GLDM**: SDLGLE | **High Value (> 0.7):** Captures small dark dependencies near bright seeds. |

### 3. 3. Shapley Additive Explanations -Based Data-Driven Radiomics Interpretation Dictionary

Figure 3 presents the class-wise SHAP contribution patterns for the 10 most influential RFs identified by the top-performing ETC. Using the model's SHAP output, we observed a clinically coherent separation between features that push predictions toward malignancy (Class 1) versus benignity (Class 0), supporting the biological mappings in our Radiological and Biological Dictionary (TU1.0).

Across the malignant class, the model primarily responds to heterogeneity, internal contrast, and outlier-bright foci, which mirror high-suspicion US patterns. The strongest discriminator is GLN from the GLRLM category (GLRLM_GLN), which strongly suppresses benign prediction (Class 0: mean SHAP = -0.1321 ± 0.1261) while supporting malignancy (Class 1: mean SHAP = +0.0460 ± 0.0413). Clinically, this aligns with a heterogeneous solid architecture, where invasive growth disrupts the uniform follicular structure, producing an uneven run-length texture. Similarly, the Intensity Range (IR) from the FO category (FO_IR) pushes toward malignancy (Class 1: +0.0554 ± 0.0484) and away from benignity (Class 0: -0.0803 ± 0.0653), consistent with nodules that contain very hypoechoic regions plus bright internal foci, creating a high-contrast internal pattern commonly seen in high-risk lesions. Outlier-sensitive intensity behavior is reinforced by Ku from the FO category which supports malignancy (Class 1: +0.0134 ± 0.0178) and opposes benignity (Class 0: -0.0506 ± 0.0577), which clinically corresponds to punctate echogenic foci/microcalcification-like signals (rare, very bright pixels on a darker background). Texture-complexity measures further support malignancy, including Dependence Variance (DV) from the GLDM category (GLDM_DV, Class 1: +0.0090 ± 0.0118, Class 0: -0.0709 ± 0.0642) and Small Dependence High Gray Level Emphasis (SDHGE) from the GLDM category (GLDM_SDHGE, Class 1: +0.0191 ± 0.0097, Class 0: -0.0353 ± 0.0062), consistent with irregular spatial organization and concentrated high-gray dependencies, which can reflect structurally complex, infiltrative tissue microarchitecture.

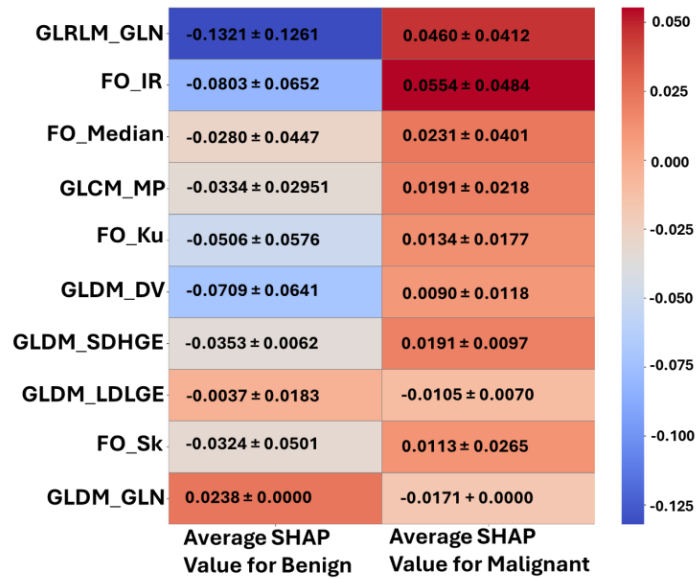

**Fig. 3.** Heatmap illustrating the average SHAP contributions of the top 10 RFs for benign (class 0) and malignant (class 1) thyroid nodules. Red regions indicate positive SHAP values (increasing the predicted likelihood of malignancy), while blue regions denote negative contributions (supporting benign classification). Features such as GLRLM_GLN, FO_IR, and GLDM_DV exhibited the most pronounced class-dependent effects, highlighting key textural and intensity-based predictors that drive the model's decision-making. SHapley Additive exPlanations (SHAP), radiomic features (RFs), First-Order (FO), Gray Level Co-occurrence Matrix (GLCM), Gray Level Run Length Matrix (GLRLM), Gray Level Dependence Matrix (GLDM), Gray Level Non-Uniformity (GLN), Dependence Variance (DV), Small Dependence High Gray Level Emphasis (SDHGE), Large Dependence Low Gray Level Emphasis (LDLGE), Intensity Range (IR), Maximum Probability (MP), Skewness (Sk), Kurtosis (Ku).

Additional features show smaller but still interpretable malignant-class tendencies. FO_Sk showed a modest positive contribution for malignancy (Class 1: +0.0113 ± 0.0266) and a negative contribution for benignity (Class 0: -0.0324 ± 0.0501), suggesting asymmetry in the intensity distribution that may reflect uneven echogenic composition. FO_Median followed a similar pattern (Class 1: +0.0231 ± 0.0401, Class 0: -0.0280 ± 0.0448), potentially corresponding to shifts in overall echogenicity and mixed tissue composition. GLCM_MP also favored malignancy (Class 1: +0.0191 ± 0.0218) over benignity (Class 0: -0.0334 ± 0.0295), indicating that repeated local gray-level relationships may still carry discriminatory value when embedded within suspicious texture patterns. Likewise, FO_Maximum (Class 1: +0.0111 ± 0.0156, Class 0: -0.0166 ± 0.0256) may reflect the presence of focal bright echoes, while GLCM_Imc2 showed a small positive contribution for malignancy (Class 1: +0.0111 ± 0.0260) and a negative contribution for benignity (Class 0: -0.0178 ± 0.0324), supporting a role for higher-order texture dependence in class separation.

A few features contributed poorly or less directionally. GLRLM_RE showed near-neutral values in both groups, although it remained slightly more positive for malignancy (Class 1: -0.0056 ± 0.0114, Class 0: -0.0067 ± 0.0275). GLSZM_ZE also displayed small negative contributions in both classes (Class 1: -0.0061 ± 0.0181, Class 0: -0.0139 ± 0.0211), suggesting a limited independent role in the top feature set. Finally, GLDM_LDLGLE showed relatively small and negative contributions for both classes (Class 1: -0.0105 ± 0.0070, Class 0: -0.0037 ± 0.0184), indicating a comparatively weak and non-directional role in class separation.

The SHAP interpretation indicates that the classifier's decision-making is anchored in clinically meaningful signals. Malignancy is driven chiefly by heterogeneity (GLRLM_GLN), high internal contrast (FO_Range), and outlier-bright foci consistent with microcalcification-like signatures (FO_Ku), with further support from dependence-based texture complexity (GLDM_DV and GLDM_SDHGLE). These findings strengthen the interpretability of the model and support the link between radiomic measurements and high-risk semantic descriptors used in TI-RADS, such as heterogeneous solid appearance, marked hypoechogenicity, and punctate echogenic foci. More details are provided in Supplemental File 1 (Sheet 6).

### 3.4 Understandable Classification Task

#### 3.4.1. Classification Results

The heatmap in Figure 4 highlights the comparative performance of the top five classifier and feature-selection combinations. Among these, the ETC combined with the SMLR Model feature selection achieved the strongest overall performance, attaining the highest Final Composite Model Score (0.960) and demonstrating consistently high values across all evaluation metrics. This combination obtained CV performance of Accuracy (0.815 ± 0.004), F1-score (0.787 ± 0.006), Precision (0.801 ± 0.006), Recall (0.815 ± 0.004), and ROC–AUC (0.772 ± 0.007). On the independent final test set, the model maintained robust performance, achieving Accuracy (0.768 ± 0.005), F1-score (0.758 ± 0.005), Precision (0.826 ± 0.004), Recall (0.768 ± 0.005), and a ROC–AUC of 0.941 ± 0.005, demonstrating strong discriminative capability and consistent generalization across datasets.

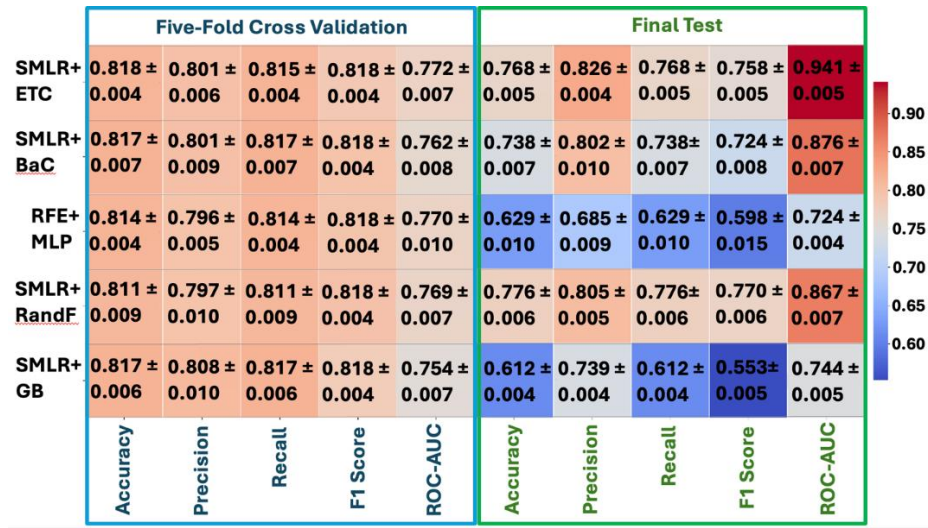

**Fig. 4.** Heatmap illustrating the performance of the top five ranked models across validation and test metrics. Each cell reports the mean value ± standard deviation for Accuracy, Precision, Recall, F1-Score, and ROC-AUC under both validation and test conditions. Rows correspond to Feature Selection Algorithms +Classifiers, and columns represent the evaluated metrics. Red shades indicate higher values, while blue shades indicate lower values. Select-From-Model with Logistic Regression (SMLR), Extra Trees Classifier (ETC), Multi-Layer Perceptron (MLP), Bagging Classifier (BaG), Recursive Feature Elimination (RFE), Random Forest (RandF), Gradient Boosting Classifier (GB), Receiver Operating Characteristic – Area Under the Curve (ROC-AUC).

Table 3 presents the selected RFs identified by the top 10 ranked models using three FSAs. Among the remaining top-performing combinations (Ranks 1–5), the BaC with SMLR ranked second with a composite score of 0.955, achieving strong final testing performance with Accuracy (0.738 ± 0.007), Precision (0.802 ± 0.010), Recall (0.738 ± 0.007), F1-score (0.724 ± 0.008), and ROC–AUC (0.876 ± 0.007). A statistically significant difference was observed between the first-ranked and second-ranked models (paired t-test, $p < 0.05$). The third-ranked MLP with RFE obtained a close score (0.954) but demonstrated reduced generalization, with lower Accuracy (0.629 ± 0.010), Precision (0.685 ± 0.009), Recall (0.629 ± 0.010), F1-score (0.598 ± 0.015), and ROC–AUC (0.724 ± 0.004). A statistically significant difference was observed between the first- and second-ranked models compared with the third-ranked model (paired t-test, $p < 0.05$), with p-values adjusted using the Benjamini–Hochberg FDR correction. The RandF with SMLR ranked fourth (score: 0.950) and maintained robust external performance, including Accuracy (0.776 ± 0.006), Precision (0.805 ± 0.005), Recall (0.776 ± 0.006), F1-score (0.770 ± 0.006), and ROC–AUC (0.867 ± 0.007). No statistically significant difference was observed between the first- and second-ranked models (paired t-test, $p > 0.05$). Boosting-based approaches showed moderate performance: GB with SMLR (rank 5) achieved Accuracy (0.612 ± 0.004), Precision (0.739 ± 0.004), Recall (0.612 ± 0.004), F1-score (0.553 ± 0.005), and ROC–AUC (0.744 ± 0.005). A statistically significant difference was observed between the fifth-ranked model compared with the first- and second-ranked models (paired t-test, $p < 0.05$), with p-values adjusted using the FDR correction. The complete performance results for all evaluated models, selected features, and hyperparameters, beyond the top ten presented here, are provided in Supplemental File 2 (Sheets 1-3).

**Table 3.** Selected features (F) from the top five models. This table presents the top 10 features selected by two different feature selection methods: SMLR (Stepwise Feature Selection by Logistic Regression) and RFE (Recursive Feature Elimination). Each method highlights a unique set of features that play a significant role in classifying thyroid nodules, aiding in the differentiation of malignant and benign nodules. First order feature (FO), Gray Level Dependence Matrix (GLDM), Gray Level Run Length Matrix (GLRLM), Gray Level Size Zone Matrix (GLSZM), Gray Level Co-occurrence Matrix (GLCM), Skewness (Sk) from the FO category (FO_Sk), Dependence Variance (DV) from the GLDM category (GLDM_DV), Run Entropy (REn) from the GLRLM category (GLRLM_REn), Maximum intensity (MaxI) from FO category (FO_MaxI), the Median from the FO category (FO_Median), Gray Level Non-Uniformity (GLN) from the GLRLM category (GLRLM_GLN), Zone Entropy (ZE) from GLSZM category (GLSZM_ZE), Kurtosis (Ku) from the FO category (FO_Ku), Informational Measure of Correlation 2 (Imc2) from GLCM category (GLCM_Imc2), Intensity Range (IR) from the FO category (FO_IR), Maximum Probability (MP) from GLCM category (GLCM_MP), Small Dependence Low Gray-Level Emphasis (SDLGLE) from the GLDM category (GLDM _SDLGLE), Large Dependence Low Gray-Level Emphasis (LDLGLE) from GLDM category.

| Method | F1 | F2 | F3 | F4 | F5 | F6 | F7 | F8 | F9 | F10 |
|---|---|---|---|---|---|---|---|---|---|---|
| **SMLR** | FO_Sk | GLDM_DV | GLRLM_REn | FO_Max | FO_Median | GLRLM_GLN | GLSZM_ZE | FO_Ku | GLCM_Imc2 | FO_IR |
| **RFE** | GLCM_MP | GLDM_DV | FO_Max | GLDM_SDLGLE | FO_Median | GLRLM_GLN | GLDM_LDLGLE | GLSZM_ZE | FO_Ku | FO_IR |

### 3.5. Interpret the Best Model Using Our Dictionaries

#### 3.5.1. Clinically Informed Interpretation of the Best Model

The SMLR + ETC, as the best-performing model in this study, leveraged a set of RFs extracted from thyroid US that capture nodule echogenicity and textural heterogeneity. These quantitative biomarkers parallel key ACR TI-RADS descriptors (e.g., echogenicity, internal composition/heterogeneity, and echogenic foci) and may therefore support malignancy risk stratification (Table 3).

**Feature 1** (FO_Sk) Quantifies asymmetry of the gray-level (echogenicity) distribution. Higher Sk can reflect an imbalanced intensity profile driven by focal hypo- or hyper-echoic components, consistent with heterogeneous nodules.

**Feature 2** (FO_DV) measures the dispersion of gray-level values. Higher value indicates greater variability in echogenicity, suggestive of complex internal texture and potentially higher-risk morphology.

**Feature 3** (Run Entropy (REn) from GLRLM category (GLRLM_REn) captures randomness/complexity of consecutive gray-level runs. Higher values indicate more irregular and heterogeneous echotexture, which may align with suspicious internal architecture.

**Feature 4** (maximum intensity (MaxI) from FO category (FO_MaxI)) represents the highest pixel intensity within the nodule. Elevated values may correspond to focal highly echogenic areas (e.g., bright interfaces or echogenic foci) that can be seen in higher TI-RADS patterns.

**Feature 5** (FO_MedI) reflects the central tendency of echogenicity. Lower median intensity (MedI) is consistent with overall hypoechogenicity, a common suspicious TI-RADS feature.

**Feature 6** (GLRLM_GLN) quantifies how unevenly gray levels are distributed. Higher values indicate greater intranodular heterogeneity, a finding frequently associated with malignant phenotypes.

**Feature 7** (Zone Entropy (ZE) from GLSZM category GLSZM_ZE) measures the complexity of homogeneous gray-level zones. Higher ZE suggests a disorganized internal structure and heterogeneous composition.

**Feature 8** (FO_Ku) describes the "peakedness" of the intensity distribution. Higher Ku implies more extreme intensity values, potentially reflecting focal bright or dark components within the nodule.

**Feature 9** (GLCM_Imc2) assesses spatial dependency and informational correlation between neighboring pixel intensities. Higher values indicate more complex spatial organization, which may be observed in structurally irregular nodules.

**Feature 10** (FO_IR) shows the difference between the minimum intensity and MaxI. A larger range reflects mixed echogenicity and a more complex internal profile.

Together, these features provide an objective translation of TI-RADS qualitative descriptors into quantitative metrics that can complement routine US-based malignancy assessment.

### 3.5.2 Shapley Additive Explanations -Based Interpretation of the Best Model

With malignancy encoded as Class 1 in best model (The SMLR + ETC), the mean SHAP effects (SHAP_Class0, SHAP_Class1) were: Feature 1 FO_Sk (+0.0108, −0.0061), Feature 2 GLDM_DV (−0.0108, +0.0061), Feature 3 GLRLM_REn (+0.0292, +0.00028), Feature 4 FO_MaxI (−0.0292, −0.00028), Feature 5 FO_MedI (+0.0140, −0.0039), Feature 6 GLRLM_GLN (−0.0140, +0.0039), Feature 7 GLSZM_ZE (+0.0090, −0.0082), Feature 8 FO_Ku (−0.0090, +0.0082), Feature 9 GLCM_Imc2 (+0.0113, −0.0042), and Feature 10 FO_IR (−0.0113, +0.0042). Under this encoding, positive SHAP values for Class 1 increase model support for malignancy, whereas negative SHAP values for Class 1 decrease malignancy support (shifting predictions toward benign/Class 0). Accordingly, Feature 2 (GLDM_DV), Feature 6 (GLRLM_GLN), Feature 8 (FO_Ku), and Feature 10 (FO_IR) showed a net positive association with malignancy prediction (positive SHAP_Class1). In contrast, Feature 1 (FO_Sk), Feature 5 (FO_MedI), Feature 7 (GLSZM_ZE), and Feature 9 (GLCM_Imc2) tended to reduce malignancy support (negative SHAP_Class1). Feature 3 (GLRLM_E) and Feature 4 (FO_MaxI) demonstrated comparatively small Class 1 contributions despite larger magnitudes in the Class 0 direction. In the following section, we provide a clinically informed interpretation of these RFs, linking them to intranodular heterogeneity and biologic correlates relevant to malignancy risk in thyroid nodules.

These RFs admit a clinically informed interpretation that directly links quantitative image properties to established US signs of malignancy and their underlying biologic correlates in thyroid nodules. Features with net positive SHAP contributions for Class 1—GLDM_DV, GLRLM_GLN, FO_Ku, and interquartile range of FO category (FO_IQR))—quantify increased intranodular heterogeneity and intensity variability. Clinically, elevated DV and GLN reflect irregular gray-level run lengths and pixel dependencies, mirroring the heterogeneous echotexture frequently observed in malignant nodules (e.g., mixed solid-cystic areas, microcalcifications, or necrosis). These changes arise from biologic hallmarks of TC, such as uneven cellular proliferation, angiogenesis, fibrosis, and psammoma body formation, all of which elevate malignancy risk under TI-RADS criteria that emphasize marked heterogeneity and punctate echogenic foci. Higher Ku indicates heavier tails in the intensity histogram (outlier bright or dark pixels), while greater IQR captures wider echogenicity spread; both correspond to the microcalcifications, cystic degeneration, or focal fibrosis commonly seen in papillary thyroid carcinoma and associated with higher Bethesda or TI-RADS scores.

Conversely, features with negative SHAP_Class1 contributions align with benign-typical US patterns. Higher FO_MedI shifts predictions toward Class 0 because malignant nodules are characteristically hypoechoic (lower gray-level values) relative to surrounding parenchyma, whereas benign nodules are more often iso- or hyperechoic. Elevated FO_Sk suggests asymmetric intensity distributions typical of benign lesions containing uniform solid areas or peripheral cystic components. Reduced GLSZM_ZE and GLCM_Imc2 indicate lower randomness in zone-size distribution and stronger pixel-value correlations, respectively; these may reflect more organized (though still irregular) texture patterns in malignancy versus the complex, less correlated textures sometimes seen in benign colloid or adenomatous nodules. Features 3 and 4 exert minimal net influence on Class 1, consistent with REn and MaxI being less discriminative once heterogeneity is accounted for. Overall, the SHAP-derived associations underscore how these RFs distill intranodular heterogeneity and echogenicity into quantifiable biomarkers with clear biologic grounding—cellular disarray, calcific deposits, and vascular irregularity—thereby providing explainable, clinically relevant support for malignancy risk stratification beyond conventional qualitative descriptors.

## 4. Discussion

The increasing detection of thyroid nodules on US has amplified the need for accurate and non-invasive methods to distinguish malignant from benign lesions. Although radiomics and ML have demonstrated considerable promise for TNC, their clinical implementation remains constrained by limited interpretability. In particular, the abstract nature of many RFs has made it difficult to relate model predictions to the semantic descriptors routinely used by radiologists. In this study, we addressed this limitation by developing a clinically anchored radiomics dictionary that maps IBSI-compliant RFs to the established semantic framework of ACR TI-RADS. By linking quantitative imaging biomarkers to clinically recognized descriptors such as composition, echogenicity, shape, margin, and echogenic foci, the proposed framework provides an interpretable bridge between computational analysis and routine thyroid US assessment.

ACR TI-RADS was developed precisely to standardize this process and reduce variability in US-based risk stratification, yet it remains fundamentally dependent on qualitative visual interpretation [20]. At the same time, radiomics and ML have shown growing promise for thyroid nodule classification, with systematic reviews and meta-analyses suggesting that US-based radiomics can achieve meaningful diagnostic performance, particularly when linked to pathology-confirmed outcomes [28]. However, the clinical translation of these models remains limited by their poor interpretability, as many RFs are mathematically abstract and difficult to reconcile with the semantic descriptors radiologists routinely use. Reviews of radiomics-TI-RADS integration have similarly emphasized that interpretability, standardization, and external validation remain major barriers to adoption [29].

A major strength of the present work is that interpretability was incorporated directly into the modeling framework rather than applied only as a post hoc explanatory step. By integrating three multicenter datasets, the study included 5,542 nodules and captured a broad range of sonographic appearances and institutional variability. Within this setting, the combination of SFLR and ETC achieved the best overall performance, with stable CV results and strong discrimination on the independent test set, including a test ROC-AUC of $0.941 \pm 0.005$. These findings indicate that a compact subset of interpretable RFs can provide robust classification performance while remaining clinically meaningful. Importantly, the selected features reflected plausible imaging correlates of malignancy, particularly intensity variation, internal heterogeneity, and focal bright outlier behavior.

The SHAP analysis further strengthened the interpretability of the proposed approach by demonstrating that the model's most influential features were aligned with established malignant and benign US patterns [28]. Features such as GLRLM_GLN, FO_IR, FO_Ku, and selected GLDM-derived metrics contributed positively to malignant classification, consistent with heterogeneous internal architecture, broader intensity dispersion, and punctate bright foci resembling suspicious TI-RADS patterns, including solid heterogeneous composition, marked hypoechogenicity, and microcalcification-like signals. In contrast, features associated with more organized and less heterogeneous texture tended to support benign classification. This concordance between SHAP-derived feature attributions and radiologically meaningful patterns suggests that the model is capturing clinically relevant image characteristics rather than spurious statistical relationships. Recent US AI studies have increasingly used explainability tools such as SHAP, but most still stop at feature ranking; our approach extends this by explicitly anchoring feature interpretation to a standardized semantic lexicon [30].

Another important contribution of this study is the adoption of a stability-aware model selection strategy. Rather than selecting models solely on the basis of average CV performance, the proposed ranking framework incorporated both predictive performance and variability across folds. This is particularly relevant in radiomics, where performance estimates may be sensitive to sampling effects, feature instability, and multicenter heterogeneity. By jointly considering discrimination and consistency, the present framework provides a more rigorous basis for identifying robust and generalizable models for thyroid US classification.

Several limitations should be considered. First, the study relied on manual segmentation, which may introduce inter-observer and intra-observer variability and affect the reproducibility of shape- and texture-based RFs. Second, the analysis was restricted to 2D static US images, which may not fully capture the three-dimensional complexity of thyroid nodules. Third, although the proposed radiomics-to-semantics mappings are clinically coherent and supported by expert review and SHAP analysis, broader validation by larger radiologist panels and correlation with histopathology or molecular markers would further strengthen their biological and clinical validity. Fourth, scanner- and acquisition-related variability remains an important challenge in US radiomics and may affect feature stability across institutions despite standardized feature extraction procedures. Finally, the RF space explored here was relatively constrained; expanding feature extraction using comprehensive libraries such as PySERA (≈500 RFs) may enable richer characterization of US patterns and strengthen both model performance and the completeness of the proposed dictionary.

Overall, the present study demonstrates that radiomics-based TNC can be made substantially more interpretable by anchoring computational features to the established ACR TI-RADS framework. The proposed dictionary and classification pipeline provide a transparent and clinically meaningful foundation for AI-assisted thyroid nodule risk stratification and support the development of more explainable and workflow-compatible decision-support tools.

## 5. Conclusion

This study presents a clinically anchored and interpretable radiomics framework for TNC in the US by directly linking quantitative imaging features to ACR TI-RADS descriptors. By aligning radiomics-based prediction with established radiological language, the proposed approach improves the transparency, consistency, and clinical relevance of ML-assisted risk stratification. The strong performance of the best-performing model, together with SHAP-based interpretation of key features related to echogenicity and internal heterogeneity, supports the potential utility of this framework as an adjunct to routine thyroid US assessment and decision-making. Further prospective multicenter validation, broader expert evaluation, and integration with automated segmentation will be important to confirm its clinical applicability and support future deployment in practice.

**Acknowledgement.** This study was supported by the Natural Sciences and Engineering Research Council of Canada (NSERC) Discovery Horizons Grant DH-2025-00119. This study was also supported by the Virtual Collaboration Group (VirCollab.com) and the Technological Virtual Collaboration (TECVICO CORP.) based in Vancouver, Canada.

**Conflict of Interest.** Authors Drs. Ali Fathi Jouzdani, Mehdi Maghsudi, and Mohammad Salmanpour were employed by the company Technological Virtual Collaboration (TECVICO Corp.). The other co-authors declare no relevant conflicts of interest or disclosures.

**Ethics Statement.** This study used only publicly available and de-identified thyroid ultrasound datasets, including TUCC, TN5000, and DDTI, which were appropriately cited in the manuscript. Since the study involved secondary analysis of anonymized public data without direct human participation or access to identifiable patient information, ethical approval and informed consent were not required.

**Code Availability.** All codes and tables are publicly shared at:
*https://github.com/MohammadRSalmanpour/Dictionary-Version-TU1.0*

# Appendix 1

**Appendix Table A1.** List of abbreviations and their definitions used in this study.

| Full Name | Abbreviation |
|---|---|
| **A** | |
| AdaBoost Classifier | ABC |
| American College of Radiology | ACR |
| American College of Radiology-Thyroid Image Reporting and Data System | ACR TI-RADS |
| Anteroposterior | AP |
| Area Under the Curve | AUC |
| Artificial Intelligence | AI |
| **B** | |
| Bagging Classifier | BaC |
| **C** | |
| Classification Algorithms | CAs |
| Cross-validation | CV |
| **D** | |
| Dependence Variance | DV |
| Dependence non-uniformity | DN |
| Digital Database Thyroid Image | DDTI |
| Dummy Classifier | DUC |
| **E** | |
| Extra Trees Classifier | ETC |
| Energy | E |
| **F** | |
| False Discovery Rate | FDR |
| Family-wise error | FWE |
| Feature Selection Algorithms | FSA |
| Fine-Needle Aspiration | FNA |
| First Order | FO |
| **G** | |
| Gaussian Process Classifier | GPC |
| Gradient Boosting | GB |
| Gray Level Co-occurrence Matrix | GLCM |
| Gray Level Dependence Matrix | GLDM |
| Gray Level Non-Uniformity | GLN |
| Gray Level Run Length Matrix | GLRLM |
| Gray Level Size Zone Matrix | GLSZM |
| **H** | |
| HistGradient Boosting | HGB |
| **I** | |
| Image Biomarker Standardization Initiative | IBSI |
| Intensity Range | IR |
| Interquartile Range | IQR |
| Informational Measure of Correlation 2 | Imc2 |
| **K** | |
| k-Nearest Neighbors | KNN |
| **L** | |
| Large Dependence High Gray-Level Emphasis | LDHGLE |
| Large Dependence Low Gray Level Emphasis | LDLGE |
| Laplacian of Gaussian | LoG |
| Linear Discriminant Analysis | LDA |
| Light Gradient Boosting Machine | LGBM |
| Low Gray-Level Run Emphasis | LGRE |
| **M** | |
| Machine Learning | ML |
| Maximum Probability | MP |
| Maximum Intensity | MaxI |
| Median Intensity | MedI |
| Multi-Layer Perceptron | MLP |
| **N** | |
| Neighboring Gray-Tone Difference Matrix | NGTDM |
| **P** | |
| Principal Component Analysis | PCA |
| **R** | |
| Radiomic Features | RFs |
| Radial Basis Function | RBF |
| Random Forest | RandF |
| Random Forest Importance | RFI |
| Recursive Feature Elimination | RFE |
| Receiver Operating Characteristic | ROC |
| Region of Interest | ROI |
| Reporting and Data Systems | RADS |
| Run Entropy | REn |
| **S** | |
| Select-From-Model with Logistic Regression | SMLR |
| Shapley Additive exPlanations | SHAP |
| Small Dependence High Gray Level Emphasis | SDHGE |
| Small Dependence Low Gray-Level Emphasis | SDLGE |
| Skewness | Sk |
| Standard Deviation | SD |
| Stochastic Gradient Descent Classifier | SGDC |
| Support Vector Machine | SVM |
| **T** | |
| Thyroid Cancer | TC |
| Thyroid Imaging Reporting and Data System | TI-RADS |
| Thyroid Nodules Classification | TNC |
| Thyroid US Cine-clip | TUCC |
| **U** | |
| Ultrasound | US |
| **V** | |
| Variance | V |
| Variance Inflation Factor | VIF |
| **W** | |
| World Health Organization | WHO |
| **X** | |
| XGBoost | XGB |
| **Z** | |
| Zone Entropy | ZE |